\documentclass[letterpaper]{article} 
\usepackage{aaai24}  
\usepackage{times}  
\usepackage{helvet}  
\usepackage{courier}  
\usepackage[hyphens]{url}  
\usepackage{graphicx} 
\usepackage{natbib}  
\usepackage{caption} 
\usepackage{algorithm}
\usepackage{algorithmic}

\usepackage{newfloat}
\usepackage{listings}
\DeclareCaptionStyle{ruled}{labelfont=normalfont,labelsep=colon,strut=off} 
\floatstyle{ruled}
\newfloat{listing}{tb}{lst}{}
\floatname{listing}{Listing}
\usepackage{array}
\usepackage{amsmath}
\usepackage{cleveref}

\usepackage{caption}
\usepackage{graphicx}
\usepackage{float} 
\usepackage{subcaption}
\usepackage{amsfonts}
\title{GanFinger: GAN-Based Fingerprint Generation for Deep Neural Network Ownership Verification}
\author {
    Huali Ren\textsuperscript{\rm 1}\textsuperscript{\rm 2},
    Anli Yan\textsuperscript{\rm 2},
    Xiaojun Ren\textsuperscript{\rm 2},
    Pei-Gen Ye,\textsuperscript{\rm 3}
    Chong-zhi Gao\textsuperscript{\rm 4},
    Zhili Zhou\textsuperscript{\rm 2},
    Jin Li\textsuperscript{\rm 2}
}
\affiliations {
   \textsuperscript{\rm 1}School of Cyberspace Security, Guangzhou University, China\\
   \textsuperscript{\rm 2}Artificial intelligence Research Institute, Guangzhou University, China\\
   \textsuperscript{\rm 3}School of Computer Science, Beijing Institute of Technology, China\\
   \textsuperscript{\rm 4}School of Computer Science, Guangzhou University, China\\
    jinli71@gmail.com
}

\usepackage{bibentry}

\begin{document}
	
	\maketitle
	
	\begin{abstract}
		Deep neural networks (DNNs) are extensively employed in a  wide range  of application  scenarios.  Generally,  training a  commercially  viable  neural  network requires  significant amounts of data and computing resources, and it is easy for unauthorized users to use the networks illegally. Therefore, network ownership verification has become one of the most crucial steps in safeguarding digital assets. To verify the ownership of networks, the existing network fingerprinting approaches perform poorly in the aspects of efficiency, stealthiness, and discriminability. To address these issues, we propose a network fingerprinting approach, named as GanFinger, to construct the network fingerprints based on the network behavior, which is characterized by network outputs of pairs of original examples and conferrable adversarial examples. Specifically, GanFinger leverages Generative Adversarial Networks (GANs) to effectively generate conferrable adversarial examples with imperceptible perturbations. These examples can exhibit identical outputs on copyrighted and pirated networks while producing different results on irrelevant networks. Moreover, to enhance the accuracy of fingerprint ownership verification, the network similarity is computed based on the accuracy-robustness distance of fingerprint examples’ outputs. To evaluate the performance of GanFinger, we construct a comprehensive benchmark consisting of 186 networks with five network structures and four popular network post-processing techniques. The benchmark  experiments demonstrate that GanFinger significantly outperforms the state-of-the-arts in efficiency,  stealthiness,  and discriminability. It  achieves a remarkable 6.57 times faster in fingerprint generation and boosts the ARUC value by 0.175, resulting in a relative improvement of about 26\%.
	\end{abstract}
	
	\section{Introduction}
	
	Deep neural network (DNN) classifiers have achieved great success in a variety of computer vision applications, such as face recognition \cite{DBLP:conf/cvpr/WangWZJGZL018}, medical image classification \cite{DBLP:journals/mia/ZhangXWX19}, and autonomous driving \cite{DBLP:journals/tits/LuoYTWF18}. Generally, training a well performing neural network is a nontrivial task \cite{press2016cleaning}, as it involves expensive data preparation (data acquisition, organization, and cleaning), massive computing resources, and expert knowledge guidance. Hence, network owners usually deploy DNN classifiers in Machine Learning as a Service (MLaaS) to share with legitimate clients for various tasks. Unfortunately, it is easy for attackers  to directly copy the network parameters and deploy the network for services without authorization \cite{DBLP:conf/uss/JagielskiCBKP20,DBLP:conf/uss/TramerZJRR16,DBLP:conf/sp/WangG18} that infringes the intellectual property rights of the network owner. As a result, to protect the intellectual property (IP) of DNNs, network ownership verification has become the first and critical step.
	
	Recently, there have been many efforts to protect the IP of DNNs, primarily through watermarking and fingerprinting techniques. Network watermarking methods either embed a pre-designed signature (e.g., a string of bits) into the parameter space of the network via specific regularization terms \cite{DBLP:conf/mir/UchidaNSS17} or embed a particular backdoor \cite{DBLP:conf/uss/AdiBCPK18} in the network to verify the identity of the suspicious network. However, these watermarking techniques interfere with the network training process and sacrifice network performance. Moreover, adaptive attacks like rewriting can swiftly eliminate the watermark \cite{DBLP:conf/sp/LukasJLK22}, making the watermarking technique invalid. In contrast to network watermarking technology, network fingerprinting methods \cite{DBLP:conf/asiaccs/CaoJG21, DBLP:conf/iscas/WangC21, li2021novel} are non-intrusive that can directly extract the unique characteristics or attributes of the owner's network as its fingerprint (i.e., unique identifier) after network distribution. As a result, network fingerprinting has emerged as a prominent alternative for IP protection of DNNs. They first construct a query set (as a fingerprint set) from the victim network, which can prove the identity of the victim network, and then they query the suspicious network with the query set. If the outputs match the victim network with a high rate, they judge that the suspicious network is a copy of the victim network.
	
	However, the existing network fingerprinting approaches suffer from three problems: (1) Low efficiency. Since they generate fingerprints from scratch when the network fingerprints need to be updated, the efficiency of fingerprint generation is relatively low; (2) Poor stealthiness. As they generate unnatural examples as fingerprints by adding considerable noise, it is relatively easy to distinguish fingerprints from original examples; (3) Limited discriminability. They only rely on the label-matching rate of fingerprint examples to attribute network ownership, ignoring the discrimination of fingerprint authenticity. Therefore, it is vulnerable to forgery attacks that use the original example as a fingerprint, leading to low-cost false accusations on unrelated networks.
	
	To cope with the challenges mentioned above, we proposed an efficient and reliable network fingerprinting framework based on Generative Adversarial Networks (GANs), named \textbf{GanFinger}. Specifically, GANs \cite{DBLP:conf/nips/GoodfellowPMXWOCB14} and two ensemble pools of positive and negative networks are first employed to generate conferrable adversarial examples with limited transferability, which is a subcategory of adversarial examples. Any conferrable example found in the victim network should have the same misclassification in a pirated network but a different one in irrelevant networks. Then, the fingerprint is constructed using a pair of (original, conferrable adversarial) examples. Ultimately, we propose a novel similarity strategy for network ownership verification, which utilizes the distance between the accuracy and robustness of the network's output on fingerprint samples to measure similarity. To verify the effectiveness of GanFinger and make a fair comparison with previous work, we built a network library containing 186 networks trained on the CIFAR10 dataset. Experimental results prove that GanFinger has superior performance in network ownership verification.
	
	In summary, our contributions are given as follows:
	\begin{itemize}
		\item The GanFinger is proposed, a novel network ownership verification framework for deep neural network (DNN) IP protection. The generator and discriminator in GanFinger enhance the efficiency and stealthiness of fingerprint generation, respectively.
		
		\item Network fingerprints are generated based on pairs of (original and conferrable adversarial) examples. The network outputs of the example pairs can ensure high discriminability of generated fingerprints.
		
		\item A network similarity measurement strategy is proposed based on accuracy-robustness distance. This strategy measures the network similarity by utilizing the accuracy inconsistency between the network outputs of original examples and the robustness consistency between the network outputs of the conferrable adversarial examples. Consequently, this strategy can measure the network similarity accurately.
		
		\item For a fair comparison, the network library was constructed to conduct experiments on 186 networks. The robustness of GanFinger is evaluated by extensive experiments on the network library, which includes five network structures with four post-processing operations. The experimental results demonstrate that compared with MetaFinger, the ARUC value of GanFinger is increased by 0.175, and the fingerprint generation time of GanFinger is about 6.57 times faster.
		
	\end{itemize}
	
	\begin{table*}[]
		\centering
		\renewcommand\arraystretch{1.1}
		\begin{tabular}{|c|c|c|c|c|c|}
			\hline
			Network structure & ResNet20 & ResNet32 & VGG16 & VGG19 & DenseNet \\ \hline
			Matching rate    & 0.868    & 0.894    & 0.896 & 0.882 & 0.890    \\ \hline
		\end{tabular}
		\caption{Label matching rate of the irrelevant network when 100 original examples are randomly sampled as fingerprint.}
		\label{table:1}
	\end{table*}
	
	\section{Related Work}
	Recent network fingerprinting works are divided into three categories according to the extracted features:  decision boundaries, neuron output states, and inter-sample relationships.
	
	(1) Fingerprinting based on decision boundaries mainly focuses on the phenomenon that the decision boundaries of the victim network and the pirate network are closer \cite{DBLP:conf/ijcai/YangWW22, DBLP:conf/cvpr/PengLCZZX22}, and characterizes the boundaries by constructing different types of adversarial examples \cite{DBLP:journals/corr/GoodfellowSS14}. \citet{DBLP:conf/asiaccs/CaoJG21} proposed the IPGuard algorithm to generate data points near the decision boundary as network fingerprints. However, IPguard is not effective under model steal attacks. Therefore, \citet{DBLP:conf/iclr/LukasZK21} exploit ensemble attacks and the transferability of adversarial examples to synthesize fingerprints called "conferrable adversarial examples" that are only transferable to pirated networks, but not to unrelated networks. Note that although we follow the idea of transferable adversarial examples, how fingerprints are generated and verified differs. 
	(2) Fingerprints based on neuron output states mainly rely on the observation that the statistical distribution of neuron outputs of the victim network and the pirated network are more similar \cite{DBLP:journals/corr/abs-2210-07481}. \citet{DBLP:conf/sp/ChenWPS0JM0S22} proposed the Deepjudge testing framework to verify network ownership from the attribute level, neuron level, and layer level. However, this approach requires the white box close to the suspect network. (3) Fingerprint-based on the relationship between samples mainly considers that the sample output of the victim network is more similar to the sample output of the pirate network. \citet{DBLP:conf/nips/GuanLH22} proposed the SAC algorithm, which selects misclassified samples as fingerprint inputs to the network and calculates the average correlation between their network outputs. However, this method requires the network output to be a posterior probability and is very sensitive to the choice of validation threshold. Although the above methods have achieved specific effects in verifying network ownership, they have yet to fully consider the efficiency, stealthiness, and authenticity of generating fingerprints in a black box scenario where only labels are output.
	
	\section{Problem Definition}
	\subsection{3.1 Threat Model}
	The network's ownership verification framework typically involves three primary actors: defenders, attackers, and trusted third parties.
	\begin{itemize}
		\item \textbf{Defender}, which is the owner of the victim network. It trains a network with good performance and deploys it in the cloud server or customer service software to provide services to authorized customers. The defender's goal is to verify a suspect network's authenticity discreetly, determine whether it is a pirated version, and subsequently present the gathered evidence to a trusted third party for further examination and evaluation.
		\item \textbf{Attacker}, whose goal is to obtain pirated networks by direct copying or network extraction. He tries to avoid piracy detection by performing post-processing methods such as fine-tuning, pruning, and input detection.
		\item \textbf{Trusted third party}, as a copyright arbiter, which receives evidence from defenders. Its goal is to verify the authenticity of fingerprints and make network ownership decisions.
	\end{itemize}
	\begin{figure*}[t]
		\centering
		\includegraphics[width=0.95\textwidth]{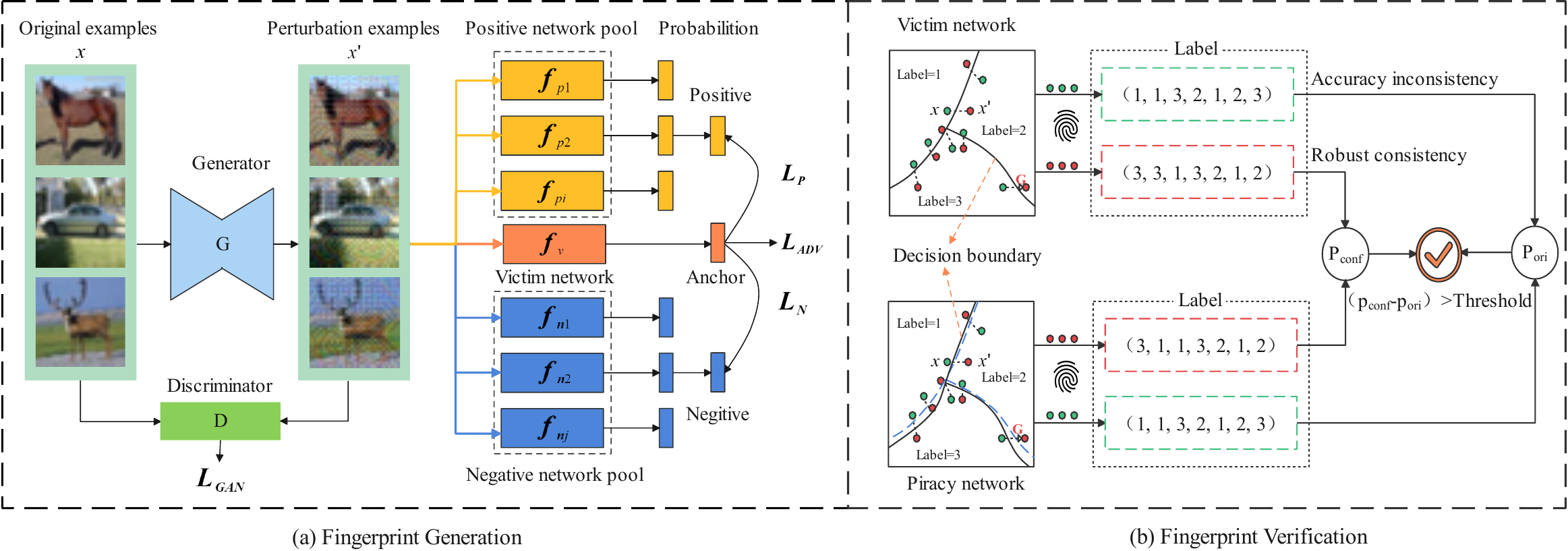} 
		\caption{The pipeline of GanFinger}
		\label{generate}
	\end{figure*}
	
	\subsection{3.2 Network Fingerprint}
	
	The scheme of fingerprint verification network ownership is generally divided into two stages:
	\begin{itemize}
		\item \textbf{Fingerprint generation}. Given white-box access to a victim network $f_{v}$, it extracts $K$ features in $f_{v}$ as network fingerprints $F_{finger}$.
		
		\item \textbf{Fingerprint verification}. Given the black-box access to the suspicious network $f_{p}$. The similarity of fingerprints $F_{finger}$ is judged through some indicators, such as label matching rate, a measure of the fraction of fingerprinting data points whose labels predicted by the suspect network match those predicted by the victim network. If the similarity is higher than the threshold, it is a pirated or otherwise irrelevant network.
		
	\end{itemize}
	However, trusted third parties need to face many different classification tasks. Malicious network owners can easily exploit this if a third party does not know the specific classification task and only verifies the network ownership through the label-matching rate. Malicious network owners may forge fingerprints to make false accusations against irrelevant networks. For example, the easiest way is to select the original samples as fingerprints directly. As shown in the table \ref{table:1}, the matching rates of irrelevant networks with different structures are over 0.86.
	
	\section{Methodology}
	
	GanFinger is an extension of the existing fingerprint-based network ownership verification, mainly consisting of three stages: network preparation, fingerprint generation, and verification. Our three phases are described in detail below.

	\subsection{4.1 Network Preparation}
	While adversarial examples can be transferable across different networks, finding conferrable adversarial examples transferable between the victim network and the pirated network but not with unrelated networks is challenging. The underlying idea of this paper is to restrict the transferability of adversarial examples by incorporating both positive, negative, and victim networks. 
	
	\textbf{Positive network pool}, i.e. pirate network. We divide the original training set into two disjoint subsets $D_{v}$ and $D_{p}$, which is the rarest case. Train the derivative network $f_{p}$ on $D_{p}$ via a label-based model extraction method.
	
	\textbf{Negative network pool}, i.e. irrelevant networks. Based on the $D_{v}$ dataset, negative network $f_{n}$ uses real labels for training, randomly initializes network parameters, and retrains from scratch.
	
	\textbf{Victim network}, i.e. source network. $D_{v}$ is used to train the victim network $f_{v}$.
	\subsection{4.2 Fingerprint Generation}
	\begin{table*}[]
		\centering
		\renewcommand\arraystretch{1.1}
		
		\begin{tabular}{|ccccc|ccccc|}
			\hline
			\multicolumn{1}{|c|}{Source(1)} & \multicolumn{4}{c|}{Fine-Tuning(20)}                                                               & \multicolumn{5}{c|}{Pruning(5)}                                                                                                    \\ \hline
			\multicolumn{1}{|c|}{ResNet20}  & \multicolumn{1}{c|}{FTLL}     & \multicolumn{1}{c|}{FTAL}  & \multicolumn{1}{c|}{RTLL}  & RTAL     & \multicolumn{1}{c|}{0.1}      & \multicolumn{1}{c|}{0.2}      & \multicolumn{1}{c|}{0.3}   & \multicolumn{1}{c|}{0.4}   & 0.5      \\ \hline
			\multicolumn{1}{|c|}{88.4}      & \multicolumn{1}{c|}{87.96}    & \multicolumn{1}{c|}{80.13} & \multicolumn{1}{c|}{87.93} & 79.86    & \multicolumn{1}{c|}{88.17}    & \multicolumn{1}{c|}{88.06}    & \multicolumn{1}{c|}{87.52} & \multicolumn{1}{c|}{85.84} & 79.8     \\ \hline
			\multicolumn{5}{|c|}{Extract-L(25)}                                                                                                  & \multicolumn{5}{c|}{Extract-P(25)}                                                                                                 \\ \hline
			\multicolumn{1}{|c|}{ResNet20}  & \multicolumn{1}{c|}{ResNet32} & \multicolumn{1}{c|}{VGG16} & \multicolumn{1}{c|}{VGG19} & DenseNet & \multicolumn{1}{c|}{ResNet20} & \multicolumn{1}{c|}{ResNet32} & \multicolumn{1}{c|}{VGG16} & \multicolumn{1}{c|}{VGG19} & DenseNet \\ \hline
			\multicolumn{1}{|c|}{87.71}     & \multicolumn{1}{c|}{88.32}    & \multicolumn{1}{c|}{89.19} & \multicolumn{1}{c|}{89.27} & 89.47    & \multicolumn{1}{c|}{88.6}     & \multicolumn{1}{c|}{88.9}     & \multicolumn{1}{c|}{89.44} & \multicolumn{1}{c|}{89.51} & 89.47    \\ \hline
			\multicolumn{5}{|c|}{Irrevelate(25)}                                                                                                 & \multicolumn{5}{c|}{Extract-Adv(25)}                                                                                               \\ \hline
			\multicolumn{1}{|c|}{ResNet20}  & \multicolumn{1}{c|}{ResNet32} & \multicolumn{1}{c|}{VGG16} & \multicolumn{1}{c|}{VGG19} & DenseNet & \multicolumn{1}{c|}{ResNet20} & \multicolumn{1}{c|}{ResNet32} & \multicolumn{1}{c|}{VGG16} & \multicolumn{1}{c|}{VGG19} & DenseNet \\ \hline
			\multicolumn{1}{|c|}{88.41}     & \multicolumn{1}{c|}{88.94}    & \multicolumn{1}{c|}{90.38} & \multicolumn{1}{c|}{90.2}  & 90.64    & \multicolumn{1}{c|}{83.23}    & \multicolumn{1}{c|}{83.73}    & \multicolumn{1}{c|}{85.65} & \multicolumn{1}{c|}{86.24} & 82.73    \\ \hline
		\end{tabular}
		
		\caption{Network library built on CIFAR10 dataset, average test accuracy(\%) of the proxy network obtained through five repeated run attacks. The number in () indicates the number of proxy networks under the attack.}
		\label{table:2}
	\end{table*}
	The fingerprint generation pipeline in GanFinger is shown in Figure \ref{generate}(a). There are mainly five modules, a generator G, a discriminator D, a victim network $f_{v}$, a positive network pool $F_{P}$ (composed of multiple positive networks $f_{pi}$), a negative network pool $F_{N}$ (composed of multiple negative networks $f_{ni}$). The generator's input is the original example $x$, the perturbation $G(x)$ is generated, and the output example $x' = x + G(x)$ after adding the perturbation. During training, we assume the white box is close to the victim and related networks in the positive and negative network pool. It is reasonable since the victim controls these networks. The loss design of the generator and discriminator is described in detail below.
	
	\textbf{Discriminator}, the purpose of discriminator D is to distinguish the perturbed data $x'$ from the original data $x$. Note that the clean data is sampled from the actual class, which is a guiding principle to encourage the generated instances to resemble the data from the original class closely. Its loss function is as follows:
	\begin{equation}
		L_{D} = E_{x}logD(x) + E_{x}log(1-D(x'))
	\end{equation}
	
	\textbf{Generator}, the purpose of the generator is to generate example $x'$  such that the discriminator cannot distinguish between real and fake examples. Meanwhile, the target and positive network output the same error label, while the output of the negative network is different.
	
	The generator expects the discriminator to output perturbed examples as real examples. Therefore, the discriminator gives the generator guidance loss as:
	\begin{equation}
		L_{GAN} = E_{x}log(1-D(x'))
	\end{equation}
	
	This paper selects FGSM \cite{DBLP:journals/corr/GoodfellowSS14} to carry out targeted attacks (target category $t$) on the target network. $L_{f_{v}}$ is the loss function for training the victim network $f_{v}$, using the cross-entropy loss. Therefore, the loss caused by cheating the target network is:
	\begin{equation}
		L_{ADV} = E_{x} L_{f_{v}}(x',t)
	\end{equation}
	
	To bind the magnitude of the perturbation, we add a soft hinge loss on the L2 norm. c is a particular perturbation upper bound formulated, conducive to training stable GAN.
	\begin{equation}
		L_{hinge} = E_{x} max(0,\left \| G(x) \right \| -c)
	\end{equation}
	
	To enhance the transferability of the generated examples, it is crucial to ensure that the output of the victim and positive network is similar while being distinct from the output of the irrelevant network. Therefore, the loss caused by conferrable adversarial examples is $L_{conf}$, where M and N are the number of networks participating in the training positive and negative network pools, and this paper takes M and N are 20. $KL$ stands for Kullback-Leibler divergence, which measures the distance between two probability distributions.
	\begin{equation}
		\begin{split}
			L_{conf} =L_P-L_N & \\= \frac{1}{M}\sum_{i=1}^{M}KL(f_{v}(x'),f_{pi}(x')) 
			& -\frac{1}{N}\sum_{j=1}^{N}KL(f_{v}(x'),f_{nj}(x'))
		\end{split}
	\end{equation}
	
	Finally, the total loss of the generator is as follows:
	\begin{equation}
		L_{loss} =  \eta L_{ADV}+ \alpha L_{GAN} + \beta L_{hinge} + \gamma L_{conf}
	\end{equation}
	
	In all our experiments, we use weights $\eta$ = 1, $\alpha$ = $\beta$ = 5, $\gamma$ = 10 and refer to Appendix C for an empirical sensitivity analysis of the hyperparameters. The discriminator and generator are trained in an alternating iterative manner. After the training, we use the generator and all networks in the validation set to generate examples with label matching rates higher than 0.9 as conferrable examples. The correctly classified original example $x$ and the corresponding misclassified conferrable adversarial example $x'$ are used to construct the fingerprint ($x$, $x'$).
	\begin{figure*}[htbp]
		\centering
		\begin{subfigure}{0.3\linewidth}
			\centering
			\includegraphics[width=0.9\linewidth]{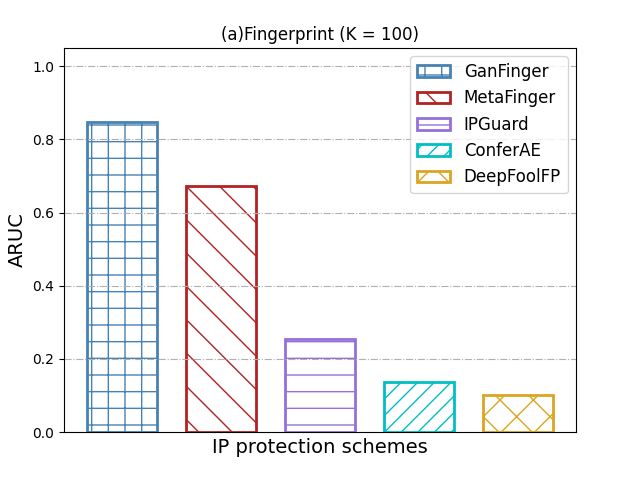}
		\end{subfigure}
		\centering
		\begin{subfigure}{0.3\linewidth}
			\centering
			\includegraphics[width=0.9\linewidth]{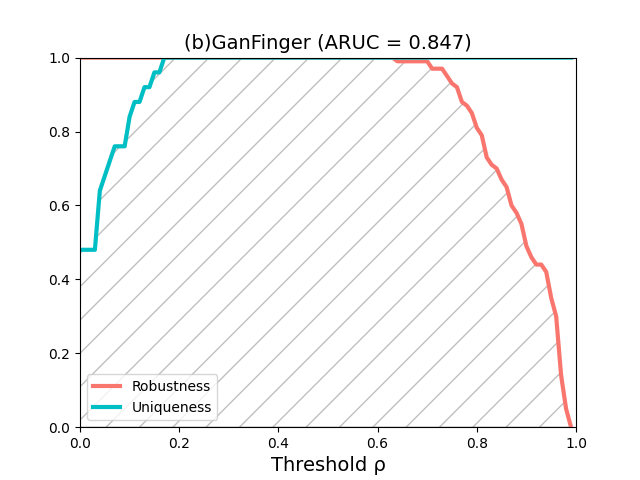}
		\end{subfigure}
		\centering
		\begin{subfigure}{0.3\linewidth}
			\centering
			\includegraphics[width=0.9\linewidth]{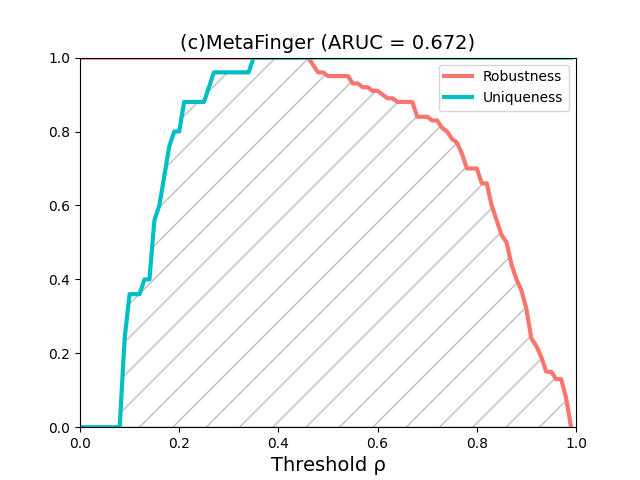}
		\end{subfigure}
		
		\centering
		\begin{subfigure}{0.3\linewidth}
			\centering
			\includegraphics[width=0.9\linewidth]{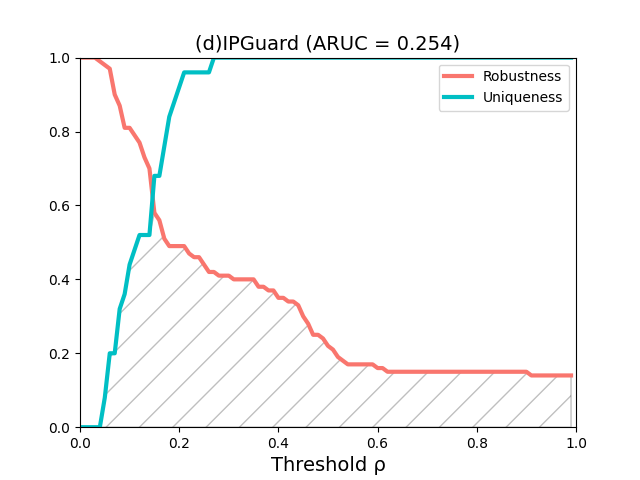}
		\end{subfigure}
		\centering
		\begin{subfigure}{0.3\linewidth}
			\centering
			\includegraphics[width=0.9\linewidth]{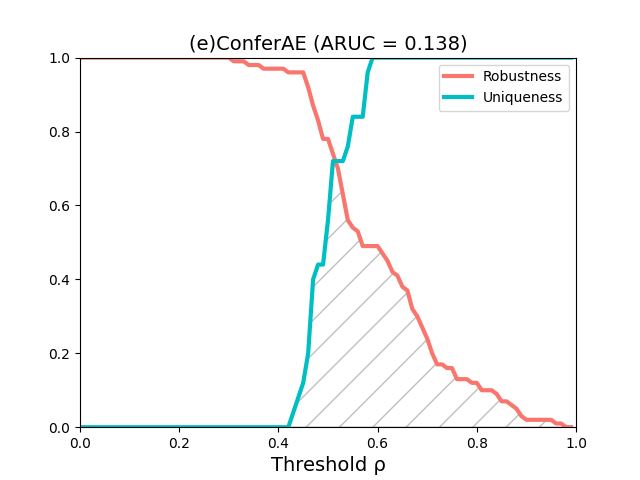}
		\end{subfigure}
		\centering
		\begin{subfigure}{0.3\linewidth}
			\centering
			\includegraphics[width=0.9\linewidth]{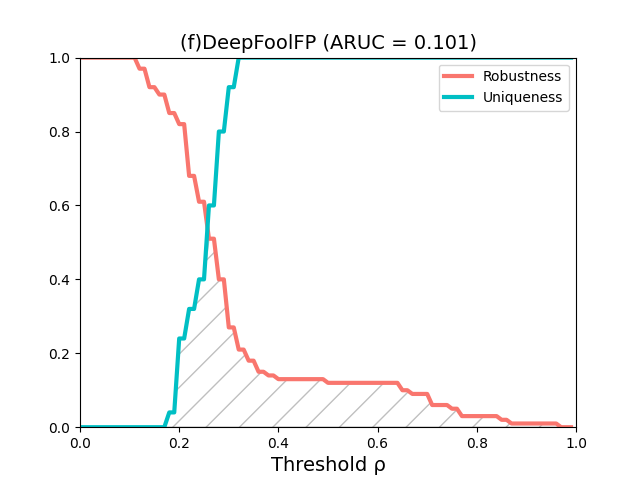}
		\end{subfigure}
		\caption{(a) ARUC values of GanFinger and baseline in the network library, and (b)-(f) change curves of robustness and uniqueness of GanFinger and baseline under different thresholds. The area of the shaded part is the size of the ARUC value.}
		\label{ARUC}
	\end{figure*}
	
	\subsection{4.3 Fingerprint Verification}
	Based on the fingerprint ($x$, $x'$) extracted in Section 4.2, we propose calculating the accuracy-robustness distance ($ARD$) to verify the copyright, and its verification framework is shown in Figure \ref{generate}(b). $ARD$ measures margin between the accuracy inconsistency $P_{ori}$ of the original example $x$ (i.e., the mismatch rate of the label) and the robust consistency $P_{conf}$ of the conferrable adversarial example $x'$ (i.e., the match rate of the label). The occurrence of different label outputs for similar examples ($x$ and $x'$) can strengthen the belief of detectors that this is a distinctive characteristic of the target network. This phenomenon highlights the unique behavior of the target network when it provides similar inputs, thereby improving the discrimination of the authenticity of fingerprints and avoiding forgery attacks. Meanwhile, $ARD$ uses the original example as a part of the fingerprint to calculate the network similarity, which can identify the network copyright more accurately.
	\begin{equation}
		ARD = P_{conf} - P_{ori}
	\end{equation}
	where,
	\begin{equation}
		P_{ori}=\frac{1}{K}\sum_{i=1}^{K}\mathbb{I}(f_{v}(x_{i})\ne f_{p}(x_{i}))
	\end{equation}
	\begin{equation}
		P_{conf}=\frac{1}{K}\sum_{i=1}^{K}[\mathbb{I}(f_{v}(x'_{i})\ne f_{v}(x_{i}))\times\mathbb{I}(f_{v}(x'_{i}) =  f_{p}(x'_{i}))]
	\end{equation}
	
	Unless otherwise specified, we follow the state-of-the-art scheme with the number K of generated fingerprints being 100. $\mathbb{I}(\cdot)$ is an indicator function that takes 1 if $\cdot$ is true, and 0 otherwise. 
	We believe that the $P_{ori}$ label mismatch rate between victims and pirated networks is lower than that of unrelated networks on the original example. The matching rate on conferrable adversarial examples $P_{conf}$ is higher than that of unrelated networks, so the ARD value of pirated networks is larger. If $ARD$ exceeds the threshold, the network to be tested is pirated and otherwise irrelevant.

	\section{Experiment}
	\subsection{5.1 Experimental Setup}
	
	To demonstrate the effectiveness of GanFinger, we separately evaluate the time-consuming, stealthy, and robustness of generating fingerprints. Meanwhile, we cover four state-of-the-art fingerprint schemes as baselines, including IPGuard \cite{DBLP:conf/asiaccs/CaoJG21}, ConferAE \cite{DBLP:conf/iclr/LukasZK21}, DeepFoolFP \cite{DBLP:conf/iscas/WangC21} and MetaFinger \cite{DBLP:conf/ijcai/YangWW22}.
	
	\textbf{Post-processing}. To avoid simple network duplication from being noticed by the owner, attackers usually modify the stolen network parameters through post-processing methods, such as fine-tuning, pruning, model extraction, and adversarial training. Details of network post-processing are deferred to Appendix A. 
	
	\textbf{Network structure}. We continue the network structure shared by most fingerprint schemes and use ResNet20 \cite{DBLP:conf/cvpr/HeZRS16}  as the victim network structure. The remaining pirated and irrelevant networks are jointly trained on neural network structures such as VGG16/VGG19 \cite{DBLP:journals/corr/SimonyanZ14a}, ResNet20/ResNet32 \cite{DBLP:conf/cvpr/HeZRS16} and DenseNet \cite{DBLP:journals/corr/IandolaMKGDK14}.
	
	\textbf{Network library}. Since the previous schemes used their network libraries for experiments, no unified benchmark existed. Therefore, this paper first constructs a library containing 186 networks in the CIFAR10 dataset. We adhere to the process of data partitioning and network building during the generation of the network (see Section 4.1). The specific use in the network library is as follows: the fingerprint generation stage includes the process of training and verification; the number of positive networks is 20 and 10, respectively, the number of negative networks is the same, and there is a victim network. Four post-processing techniques are used to generate the pirate network in the fingerprint verification stage. We tested five different networks for each setting to avoid the chance of random weights. The details of its structure, performance, and quantity are shown in Table \ref{table:2}.
	\begin{table*}[]
		\centering
		\renewcommand\arraystretch{1.1}
		\begin{tabular}{|c|c|c|c|c|c|}
			\hline
			Scheme       & IPGuard & ConferAE & MetaFinger & DeepFoolFP & GanFinger \\ \hline
			Generation & 163.21s & 56,668.16s                & 322.49s    & 221.79s    & 49.06s    \\ \hline
			Verification & 2.08s   & 2.37s                      & 2.42s      & 2.25s      & 2.71s     \\ \hline
		\end{tabular}
		\caption{Comparing the time consumption of generating and verifying fingerprints.}
		\label{table:3}
	\end{table*}
	
	\textbf{Evaluation metrics}. We employ two metrics to verify the effectiveness of the fingerprinting scheme: accuracy-robustness distance (ARD) and area under the uniqueness curve (ARUC). The larger their values, the better the performance of the scheme. Please refer to Appendix B for detailed definitions and calculation procedures.
	
	\textbf{Implementation details}. We train a conditional adversarial network to generate conferrable adversarial examples. Similar to AdvGAN \cite{DBLP:conf/ijcai/XiaoLZHLS18}, the generator and discriminator structures in GanFinger are composed of multi-layer neural networks. We employ the FGSM approach for a targeted attack on the victim network. The adversarial loss is computed using the CW loss function to limit the size of the generated perturbation L to 0.05. GanFinger utilizes a batch size of 128, a learning rate of 0.001, and training for 60 epochs with the SGD optimizer. 
	
	\subsection{5.2 Effectiveness Analysis}
	To demonstrate the effectiveness of GanFinger, we first perform a unified comparison with state-of-the-art fingerprinting methods on the network library. The overall result is shown in Figure \ref{ARUC}. From the Figure \ref{ARUC}(a), it can be seen that the maximum ARUC obtained by GanFinger is 0.847, which is 0.175 higher than that of MetaFinger, and the relative improvement is about 26\%. The larger the ARUC value, the stronger the ability to distinguish pirated networks from irrelevant networks. Figure \ref{ARUC}(b)-(f) shows the robustness and uniqueness curves of each fingerprint scheme.
	\begin{figure}[t]
		\centering
		\includegraphics[width=0.4\textwidth]{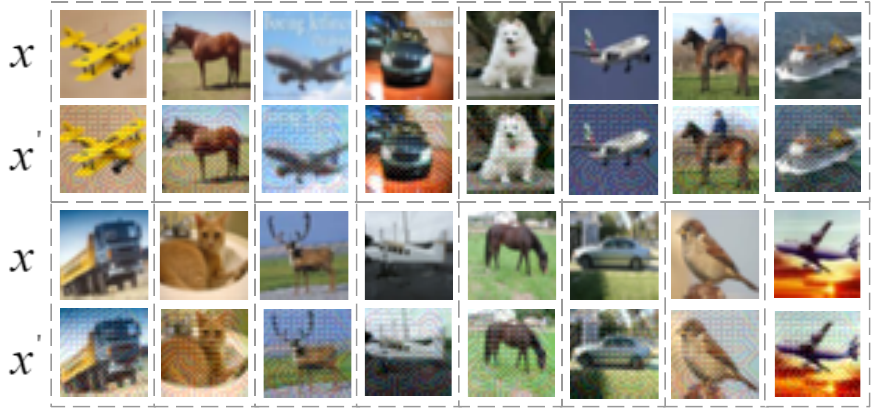} 
		\caption{Fingerprint ($x,x'$) constructed by GanFinger. Among them, the row of $x$ represents original example, the row of $x'$ represents the generated conferrable adversarial example.}
		\label{imperceptibility}
	\end{figure}
	\begin{figure}[t]
		\centering
		\includegraphics[width=0.4\textwidth]{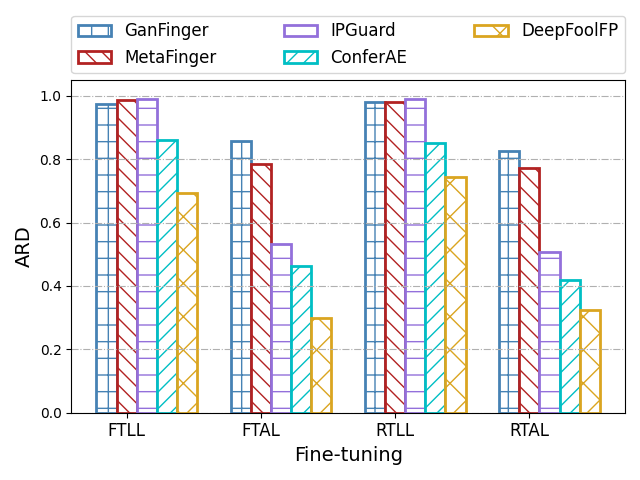} 
		\caption{ARD value under fine-tuning attack}
		\label{fineturn}
	\end{figure}
	\begin{figure}[t]
		\centering
		\includegraphics[width=0.4\textwidth]{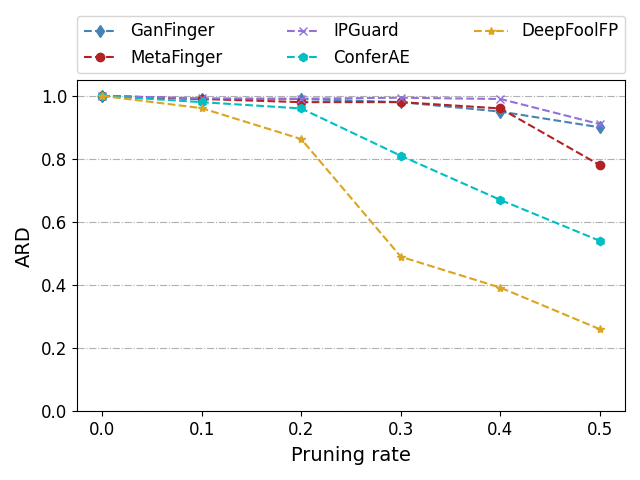} 
		\caption{ARD value under pruning attack}
		\label{pruning}
	\end{figure}
	\begin{figure*}[htbp]
		\centering
		\begin{subfigure}{0.3\linewidth}
			\centering
			\includegraphics[width=0.9\linewidth]{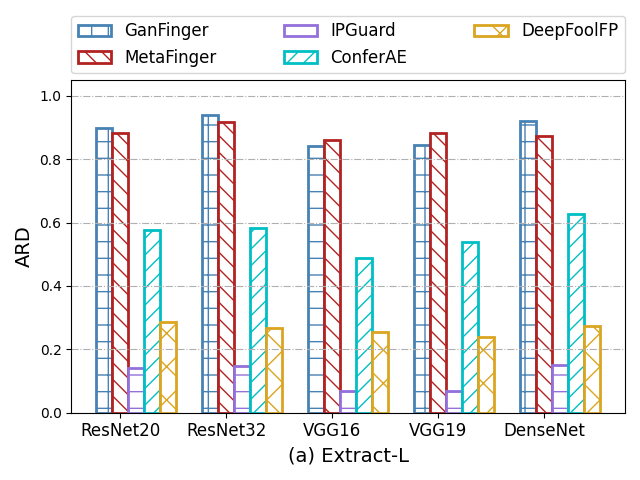}
		\end{subfigure}
		\centering
		\begin{subfigure}{0.3\linewidth}
			\centering
			\includegraphics[width=0.9\linewidth]{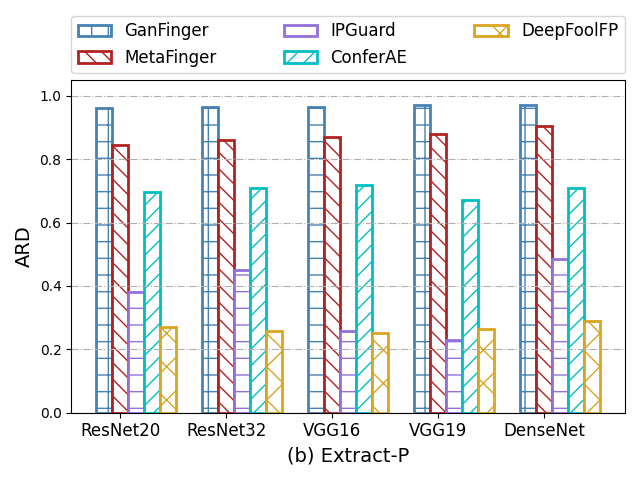}
		\end{subfigure}
		\centering
		\begin{subfigure}{0.3\linewidth}
			\centering
			\includegraphics[width=0.9\linewidth]{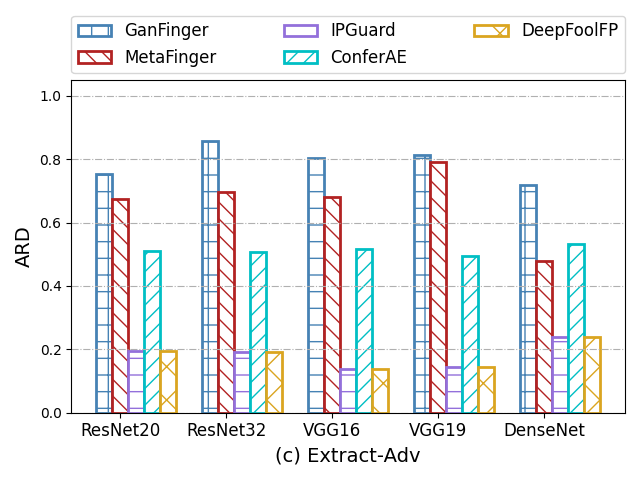}
		\end{subfigure}
		\caption{ARD value of extraction and adversarial training under different network structures, (a) output label-based extraction attack, (b) posterior probability-based extraction attack, (c) adversarial training}
		\label{extraction}
	\end{figure*}
	\subsection{5.3 Time-consuming}
	To evaluate the efficiency of GanFinger, we need to consider the time overhead required for fingerprint generation and verification separately. It is worth noting that the network preparation time was not taken into account since it is a shared process. The experimental results are shown in Table \ref{table:3}. Regarding fingerprint generation time, our method is 6.57x and 3.32x faster than the best-performing MetaFinger and the most efficient IPGuard. In terms of verification time, GanFinger is close to the baseline scheme.This demonstrates that the ARD metric can accomplish the verification task without sacrificing efficiency.

	\subsection{5.4 Stealthiness}
	To ensure the normality and repeatability of the forensics process, the stealthiness of fingerprints is an indicator that must be considered. Because attackers may provide other wrong labels or directly refuse to serve examples that perceive anomalies to evade detection. We visualize the constructed fingerprints as shown in Figure \ref{imperceptibility}. It can be found that the fingerprints generated by GanFinger have good visual consistency. At the same time, for third-party detectors, even for examples of unknown categories, the respective outputs of the pirate network $f_p$ and the victim network $f_v$  on the fingerprint example ($x$, $x'$) satisfy ($f_p(x) = f_v(x)$) $\ne$ ($f_p(x')=f_v(x')$). This unique behavioral feature can make the detection results more reliable.
	
	\subsection{5.5 Robustness}
	
	To evade piracy detection mechanisms, attackers often use post-processing operations such as fine-tuning, pruning, and adversarial training to modify stolen network weight parameters before the network is made public.
	
	\textbf{Fine-tuning.} As shown in Figure \ref{fineturn}, We first validated the robustness of different fine-tuning methods on GanFinger and baseline using the accuracy-robustness distance (ARD) metric. When fine-tuning the last layer of FTLL and RTLL, we observe that the ARD value of fingerprint pairs constructed by GanFinger reaches over 97\%. When fine-tuning all layers in FTAL and RTAL, GanFinger outperforms other schemes with an ARD value higher than 82\%.
	
	\textbf{Pruning.} 
	We prune at a rate of p\% (set the p\% parameter with the smallest weight value in each layer to 0). Table \ref*{table:2} shows that when the pruning rate is 50\%, the network's performance has been reduced by about 9\%. If pruning continues, the performance of the network will drop significantly. Therefore, the maximum pruning rate of the pirated network we tested is 50\%. It can be seen from Figure \ref{pruning} that as the pruning rate increases, the ARD values of all schemes show a downward trend. Especially ConferAE and DeepFoolFP, their ARD values drop sharply. But the ARD value of GanFinger remains above 90\%.
	
	\textbf{Model extraction.} When the attacker approaches the black-box victim network, based on the amount of information output by the victim, this paper mainly tests the robustness of fingerprints under extraction attacks where the output is the label (Extract-L, less information) and the posterior probability (Extract-P, more information). The experimental results are shown in Figures \ref{extraction} (a) and (b). On the whole, Extract-P has a higher ARD value than Extract-L in various ownership verification schemes, which means that Extract-P is more likely to detect pirated networks than Extract-L successfully. We believe this phenomenon is reasonable because the posterior probability contains more information about the victim network than the label, resulting in the pirate network based on the posterior probability of stealing being more similar to the victim network.
	
	When both the attacker and the victim are ResNet20 networks, GanFinger achieves 90\% and 96\% ARD values in Extract-L and Extract-P better than the baseline, which shows that the fingerprints generated by GanFinger are robust to model extraction attacks. In addition, both GanFinger and the schemes in the baseline achieve relatively stable ARD when the attacker adopts different network structures, which shows that GanFinger is robust to the structure types of model extraction attacks.
	\begin{table}[]
		\centering
		\renewcommand\arraystretch{1.1}
		\begin{tabular}{ccccc}
			\hline
			\multicolumn{1}{c|}{Network pool size} & \multicolumn{1}{c|}{5}      & \multicolumn{1}{c|}{10}     & \multicolumn{1}{c|}{15}     & 20                   \\ \hline
			\multicolumn{1}{c|}{ARUC}         & \multicolumn{1}{c|}{0.6323} & \multicolumn{1}{c|}{0.7257} & \multicolumn{1}{c|}{0.7906} & 0.8470                \\ \hline
			\multicolumn{1}{l}{}              & \multicolumn{1}{l}{}        & \multicolumn{1}{l}{}        & \multicolumn{1}{l}{}        & \multicolumn{1}{l}{}
		\end{tabular}
		\caption{ARUC value under different network numbers during training.}
		\label{table:4}
	\end{table}
	
	\textbf{Adversarial training.} To avoid the copyright detection mechanism, attackers often use adversarial training to improve the robustness of the piracy network. See Appendix A for details. The ARD value of the network ownership verification scheme under different network structures during adversarial training is shown in Figure \ref{extraction}(c). We observe that GanFinger is higher than the baseline in other structures. In particular, GanFinger achieved the highest ARD of 85.8\% on ResNet32, and the lowest was 72\% on DenseNet. In the baseline scheme, DeepFooFP has the lowest ARD of 13.7\%.
	
	\begin{table}[]
		\centering
		\renewcommand\arraystretch{1.1}
		\begin{tabular}{ccccc}
			\hline
			\multicolumn{1}{c|}{G loss} & \multicolumn{1}{c|}{$\eta$}    & \multicolumn{1}{c|}{$\alpha$}  & \multicolumn{1}{c|}{$\beta$}   & $\gamma$                \\ \hline
			\multicolumn{1}{c|}{ARUC}   & \multicolumn{1}{c|}{0.8102} & \multicolumn{1}{c|}{0.7361} & \multicolumn{1}{c|}{0.8025} & 0.5324                \\ \hline
			\multicolumn{1}{l}{}        & \multicolumn{1}{l}{}        & \multicolumn{1}{l}{}        & \multicolumn{1}{l}{}        & \multicolumn{1}{l}{}
		\end{tabular}
		\caption{The ARUC value when $\eta$, $\alpha$, $\beta$, and $\gamma$ in the generator loss are 0, respectively.}
		\label{table:5}
		
	\end{table}
	\subsection{5.6 Ablation Experiment}
	In the ablation study, we quantitatively analyze the impact of network pool size and various loss parts of the generator on GanFinger during training.
	
	\textbf{Effect of network pool size.} As shown in table \ref{table:4}. It can be found that as the number of networks in the network pool increases during training, the ARUC value gradually increases. This suggests that increasing the number of networks in the network pool during training can lead to greater diversity and enhance the robustness of the generated fingerprints.
	
	\textbf{Effect of each loss term in the generator.} As shown in Table \ref{table:5}. It can be found that when $\eta$, $\alpha$, $\beta$, and $\gamma$ are respectively 0, the corresponding ARUC is less than 0.847 when combined. This shows the effectiveness of our various parts. In addition, we analyzed the sensitivity of $\eta$, $\alpha$, $\beta$, and $\gamma$ values in Appendix C, indicating the effectiveness of GanFInger
	
	\section{Conclusion}
	This paper specifically addresses the issue of neural network ownership verification in the context of fingerprinting. The proposed method GanFinger aims to generate stealthy, robust, and authentic fingerprints efficiently. GanFinger exploits unique output patterns exhibited by the network on fingerprints (consisting of pairs of original and conferrable adversarial examples) to characterize ownership. It employs transfer techniques to generate conferrable adversarial examples that are consistent on victim and pirate networks but differ on unrelated networks. At the same time, an accuracy-robustness distance is proposed to calculate the similarity of the network for ownership verification. Experimental results on the network library show that our method significantly and consistently outperforms the state-of-the-art methods.
	
	\bibliography{aaai24v3}

	\section*{Appendix}
	\appendix
	\section{A. Post-processing}
	
	\begin{itemize}
		\item \textbf{Fine-tuning:} There are four ways to fine-tune the network. Fine-Tune Last Layer (FTLL): fine-tune the weight of the last layer of the network and freeze the rest of the network; Fine-Tune All Layers (FTAL): fine-tune the weights of all layers of the network; Re-Train Last Layer (RTLL): Randomly initialize the last layer of the network and fine-tune the weights of the last layer; Re-Train All Layers (RTAL): Randomly initialize the last layer of the network and fine-tune the weights of all layers. We assume the network is fine-tuned in the experiments using the attacker dataset $D_p$ and iteratively trained ten times.
		
		\item \textbf{Pruning:} Weight pruning is a standard network compression method to remove redundant weights. In the experiment, we reset the weight with the smallest absolute value to 0 and increased the pruning ratio $p$ from 0.1 with a step of 0.1. Considering the impact of pruning on network performance, we select networks with a pruning rate of 0.1 to 0.5 and put them into the positive network pool.
		\item \textbf{Model extraction:} When the attacker can only access the API provided by the victim network (black box) in the cloud server, the attacker can use two types of model extraction attacks based on posterior probability and labels. In label-based extraction, the attacker uses the label output by the victim network as the example label, calculates the cross-entropy between the two as the loss function, and trains a local surrogate network. In the model extraction based on posterior probability, the KL divergence value between the target network and the output probability distribution of the extracted network is calculated as the loss function. For the detailed algorithm, please refer to SAC (Guan, Liang, and He 2022). In the experiment, we assume that the attacker dataset $D_p$ is used to query the victim network, to train a local piracy network.
		\item \textbf{Adversarial training:} which is the standard approach for adversarial example defense. In the existing piracy network checking, most of the work is based on the mechanism of adversarial examples to generate fingerprints. Therefore, to avoid inspection, the attacker may add adversarial examples to the training of the pirated network to improve its robustness. In the experiment, we assume that based on the label extraction network, the FGSM method generates adversarial examples on the attacker dataset $D_p$. Then two rounds of adversarial training are performed to generate a pirated network.
	\end{itemize}
	
	\section{B. Evaluation metrics} 
	\begin{itemize}
		\item \textbf{Accuracy-robustness distance (ARD):} The ARD value is used to measure the similarity of the victim and suspicious networks' predicted behavior to the fingerprint pair. The specific calculation is in Section 4.3. This paper's accuracy-robustness distance corresponds to the matching rate in other fingerprinting schemes. 
		\item \textbf{Area under the uniqueness curve (ARUC):} The ARUC value measures the area of the intersection region under Robustness and Uniqueness when the threshold value varies between (0, 1). It can be used to ignore the influence of different thresholds for overall evaluation. See IPGuard (Cao, Jia, and Gong 2021) for detailed calculations. Robustness refers to the proportion of genuine pirated networks identified as pirated (i.e., true positive rate). Uniqueness refers to the proportion of true irrelevant networks identified as irrelevant (i.e., true negative rate).
	\end{itemize}
	
	\begin{table}[]
		\centering
		\resizebox{0.45\textwidth}{!}{
			\renewcommand\arraystretch{1.1}
			\begin{tabular}{|c|c|c|c|c|c|c|c|c|l|}
				\hline
				$\eta$ & $\alpha$ & $\beta$ & $\gamma$ & ARUC   & $\eta$ & $\alpha$ & $\beta$ & $\gamma$ & \multicolumn{1}{c|}{ARUC} \\ \hline
				1      & 5        & 5       & 1        & 0.6843 & 0.1    & 5        & 5       & 1        & 0.7981                    \\ \hline
				1      & 5        & 5       & 10       & 0.8474 & 0.1    & 5        & 5       & 10       & 0.8223                    \\ \hline
				1      & 5        & 0.5     & 1        & 0.6969 & 0.1    & 5        & 0.5     & 1        & 0.8205                    \\ \hline
				1      & 5        & 0.5     & 10       & 0.7336 & 0.1    & 5        & 0.5     & 10       & 0.8264                    \\ \hline
				1      & 0.5      & 0.5     & 1        & 0.6748 & 0.1    & 0.5      & 0.5     & 1        & 0.7303                    \\ \hline
				1      & 0.5      & 0.5     & 10       & 0.7772 & 0.1    & 0.5      & 0.5     & 10       & 0.8209                    \\ \hline
				1      & 0.5      & 5       & 1        & 0.6539 & 0.1    & 0.5      & 5       & 1        & 0.8073                    \\ \hline
				1      & 0.5      & 5       & 10       & 0.8311 & 0.1    & 0.5      & 5       & 10       & 0.8375                    \\ \hline
		\end{tabular}}
		\caption{An empirical sensitivity analysis of the hyperparameters in Eq. 6.}
		\label{table:6}
	\end{table}
	
	\section{C. Empirical sensitivity analysis} 
	
	We perform a grid search to understand the sensitivity of hyperparameter selection in Equation 6. Evaluate the area under the uniqueness curve (ARUC) scores for different combinations of $\eta$, $\alpha$, $\beta$, $\gamma$ for n = 100 inputs while keeping the remaining parameters constant on the model library. We combined the optimal values 1, 5, 5, and 1 of $\eta$, $\alpha$, $\beta$, and $\gamma$ with 0.1, 0.5, 0.5 and 1 after being reduced by 10 times, for a total of $2^4$ groups of experiments.
	
	The results are shown in Table \ref{table:6}. We found that the ARUC value will change when $\eta$, $\alpha$, $\beta$, and $\gamma$ take different values, but we still have 93\% of the combinations with an ARUC value higher than the comparison work (ARUC is 0.672). In addition, when the value of $\gamma$ is small, it has a greater impact on the performance of GanFinger. When $\gamma$ is large relative to other parameters, we measure higher ARUC values.

\end{document}